# Electrically driven quantum light emission in electromechanically-tuneable photonic crystal cavities

M. Petruzzella*,[1, 1] F.M. Pagliano, Z. Zobenica, S. Birindelli, M. Cotrufo, F.W.M van Otten, R.W. van der Heijden, and A. Fiore[2]

*Department of Applied Physics, Eindhoven University of Technology, P.O. Box 513, NL-5600MB Eindhoven, The Netherlands*

(Dated: 08 October 2017)

[1] Electronic mail: m.petruzzella@tue.nl

[2] Electronic mail: a.fiore@tue.nl

**A single quantum dot deterministically coupled to a photonic crystal environment constitutes an indispensable elementary unit to both generate and manipulate single-photons in next-generation quantum photonic circuits. To date, the scaling of the number of these quantum nodes on a fully-integrated chip has been prevented by the use of optical pumping strategies that require a bulky off-chip laser along with the lack of methods to control the energies of nano-cavities and emitters. Here, we concurrently overcome these limitations by demonstrating electrical injection of single excitonic lines within a nano-electro-mechanically tuneable photonic crystal cavity. When an electrically-driven dot line is brought into resonance with a photonic crystal mode, its emission rate is enhanced. Anti-bunching experiments reveal the quantum nature of these on-demand sources emitting in the telecom range. These results represent an important step forward in the realization of integrated quantum optics experiments featuring multiple electrically-triggered Purcell-enhanced single-photon sources embedded in a reconfigurable semiconductor architecture.**

**INTRODUCTION**

The generation of a pure and deterministic stream of quantum states of light is at the heart of future quantum photonic technologies [1] which promise physically-secure communication networks, quantum-enhanced sensing and quantum simulators able to emulate the dynamics of classically-intractable many-body Hamiltonians [2, 3]. Quantum Dots (QDs), realized by growing semiconductor nano-islands into a hosting material of wider bandgap, have emerged as the ultimate on-demand sources of single and entangled photons [4]. In fact, the single-photon emission from resonantly-excited QDs currently competes in terms of indistinguishability and purity with heralded sources based on nonlinear processes, while providing a single-photon brightness one order of magnitude higher [5]. Importantly, the ability to monolithically incorporate these emitters into III-V materials opened the possibility to build active quantum photonic integrated circuits (QPICs) where the manipulation of path-encoded flying qubits can be implemented by an integrated linear-optics platform composed of ridge waveguides, beam splitters and phase shifters [6]. In this context nano-cavities, such as photonic crystals (PhCs), represent an attractive toolbox to enhance the exciton-photon interaction by exploiting their low-loss optical modes confined in a sub-wavelength region.

In fact, these have been proposed both to significantly boost the single-photon generation rate through the Purcell effect [7] and to enable nonlinear interactions at the single-photon level [8, 9]

So far, optical pumping has been adopted as the only excitation method to study cavity quantum electrodynamics (cQED) phenomena originating from the coupling of quantum emitters to photonic crystal structures. However, in the context of QPIC, this poses severe practical limitations in realizing photonic circuits comprising multiple sources. Firstly, the stability of the external pump laser and of its free-space optical path limits the maximum integration time employed in correlation experiments. Secondly, the excitation area that can be addressed optically is limited by the numerical aperture of the objective (usually in the order of tens of square microns), which restrains the number of sources that can be independently excited. Lastly and most importantly, the scattering of the pump photons into integrated single-photons detectors [10] originates stray counts that affect the visibility of experiments with sources and detectors integrated on the same chip [11, 12].

Electrical injection represents a handy solution to overcome these drawbacks. By sandwiching the QD-layer in a p-i-n junction, single-photon emitting diodes (SPED) have been successfully demonstrated [13] and adopted in a number of free-space quantum protocols which include two-photon interference [14, 15], entanglement [16, 17, 18], teleportation [19, 20], quantum key distribution [21, 22] and quantum relays [23]. Electrical injection methods have been implemented also within vertical cavities, such as planar Bragg micro-cavities [24] and nanopillars [25] in order to improve the out-of-plane single-photon efficiency. However, these structures are not suited for funnelling single-photons to on-chip waveguides, contrary to PhCs modes that can be engineered to efficiently couple to an in-plane integrated circuitry. Despite several demonstrations of electrical injection in PhC cavities [26, 27, 28, 29] emission from single excitons in electroluminescence has not been demonstrated so far [30].

Additionally, due to the inhomogeneous broadening of semiconductor emitters and fabrication imperfections, an energy mismatch between QDs and cavity resonances is unavoidable. Consequently, local and reversible methods to bring multiple QDs and photonic modes into a common resonance are crucial to produce indistinguishable single-photons and to scale the number of distinct integrated cavity-emitter systems on a chip.

In this work, we report two major steps towards the development of electrically-injected on-chip single-photon sources: 1) we demonstrate electroluminescence from single QDs coupled to a PhC cavity and the antibunched nature of the emitted light and 2) we integrate electromechanical tunability within this PhC LED structure. By combining the nano-electro-mechanical actuation of a double-layer PhC membrane with the electrical excitation of QDs, a single exciton is brought on resonance with a tuneable cavity mode, and its emission is enhanced. This will enable fully-tuneable, electrically-injected cavity-enhanced single-photon sources.

**MATERIALS AND METHODS**

The core of the device reported here consists of a pair of parallel photonic crystal slabs (Fig.1a) [31, 32]. When their distance (d) is sufficiently small to enable evanescent coupling, their individual resonances split in a pair of super-modes spreading over the two membranes, which have a symmetric (S) and anti-symmetric (AS) profile along the growth direction. An electrostatic force, originated by the voltage applied between the membranes, controls the physical separation between the slabs, inducing a reversible change in the mode coupling. This results in an opposite wavelength shift of the S and AS mode. The QD-layer, located in the top membrane, can be excited independently by injecting a current through a p-i-n diode realized across the top membrane.
The sample is grown by molecular beam epitaxy and includes two 170 nm-thick GaAs slabs separated by a 200nm-thick $Al_{0.7}Ga_{0.3}As$ sacrificial layer. At the centre of the top slab a layer of self-assembled InAs quantum dots is grown. Their ground state emission is centred at 1225 nm at low temperature (9 K). In order to realize the two p-i-n diodes, for cavity actuation and QD-injection, the top 50-nm-thick part of the two membranes is p doped, while the bottom 50-nm thick part of the top membrane is n doped ($p_{QD}=1.5\times10^{18}$, $n=p_{CAV}=2\times10^{18}cm^{3}$). The fabrication of the device consists in two main parts (A and B). In part A, the mesa-structure and the contact layers of the two diodes are fabricated. This process comprises three sequential optical lithography steps (i-iii). (i) The n-mesa of both diodes is lithographically defined and etched via a citric acid solution. The time and the concentration of this solution is calibrated to reach the n-doped layer and

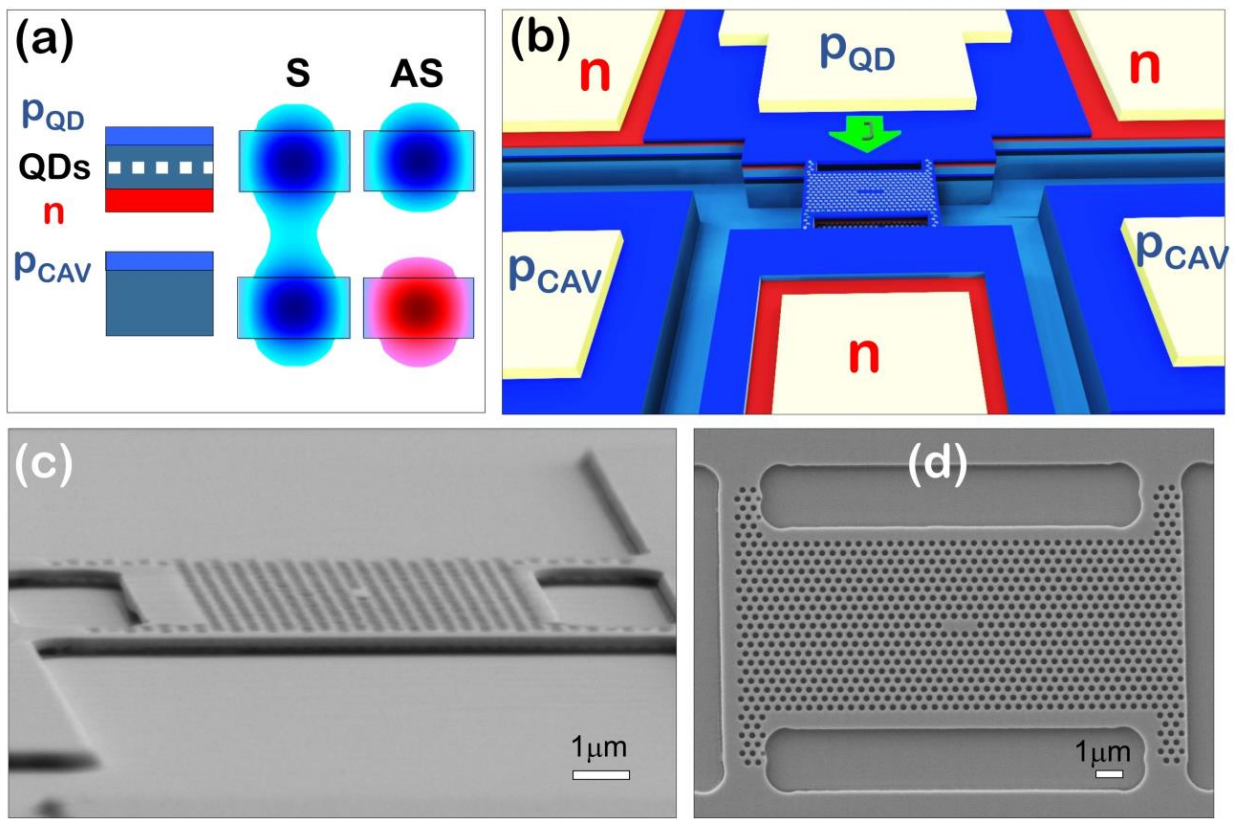

**Figure 1** (a) Sketch of the symmetric (S) and anti-symmetric (AS) modes originating by the vertical coupling between the two slabs along with the vertical layers of the device. (b) Sketch of the in-plane contact layout of the device, showing the two diodes surrounding the cavity-bridge region for current injection and cavity actuation. (c) Tilted scanning electron micrograph of the device where the top free-standing membrane is visible and is connected to a supporting frame via four micro-arms. (d) Top-view of the PhC micro-bridge embodying an L3 cavity at its centre the final etching profile is measured by a profilometer. (ii) The p-mesa of the cavity diode is created by the combination of dry etching based on a $SiCl_4$ recipe and wet etching in a hydrofluoric acid solution to selectively remove the inter-membrane AlGaAs. In this step the micro-bridge employed for the cavity actuation is also defined. (iii) The contact layers (Ti/Au alloy, 50/200 nm) are deposited on all the three doped surfaces through electron beam evaporation and lift-off. In part B, a 400nm-thick Silicon Nitride ($Si_3N_4$) layer is deposited via plasma enhanced deposition technologies. This film acts both as a hard mask for the definition the photonic crystal holes and as a conformal coating which increases the stiffness of the micro-bridge, reducing the probability of collapsing during the process. The photonic crystal layout is aligned to the centre of the bridge geometry, patterned by electron beam lithography and transferred to the $Si_3N_4$ and then to both membranes by a deep dry etching step using a high-temperature $Cl_2/N_2$ recipe. Finally, the two AlGaAs layers are removed by selective wet etching (HCl solution) and and super-critical drying is employed to avoid stiction arising from capillary forces. At the end of the process, the SiN mask is removed isotropically employing a dry etching based on $CF_4$. Notice that although both membranes are released during the fabrication, the top membrane is much more compliant than the bottom one, since the latter is clamped from all sides.

In the following experiments the employed in-plane PhC design is composed of a triangular lattice of air holes where a point-defect cavity is realized rearranging the positions of four holes around a central point (H0) or removing three (L3) holes (Fig.1d). The lattice constant is designed to match the fundamental PhC mode with the QD ground state emission.

A key challenge for the electrical pumping of PhC cavities is the injection of both electrons and holes in the cavity region, while keeping the optical loss low. This is achieved in this work by placing the p and n contacts of the QD diode on the opposite sites of the cavity which forces the current to flow laterally along the micro-bridge. Fig.1b shows a sketch of the device. The two distinct series of contacts for carrier injection and cavity tuning are realized using the configuration [n, $p_{QD}$, n] and [$p_{CAV}$, n, $p_{CAV}$].

Low-temperature (T=9 K) micro-electroluminescence ($\mu$EL) measurements are carried out in a continuous-flow He cryostat fitted with two electrical probes set at a common ground. The emitted $\mu$EL is collected through an objective (numerical aperture NA=0.4) and analysed in a fiber-coupled spectrometer. While in the experiments reported here µEL is collected from the top of the PhC, we have already shown that the PhC emission can be efficiently funnelled to ridge waveguides [33, 34].

**RESULTS AND DISCUSSION**

In a first set of experiments we study the tunability of the PhC LED. To this aim, the QD ensemble is electrically excited with a relatively high current (~ mA), while the cavity wavelength is varied by changing the vertical position of the top membrane by an electrostatic voltage ($V_{CAV}$).

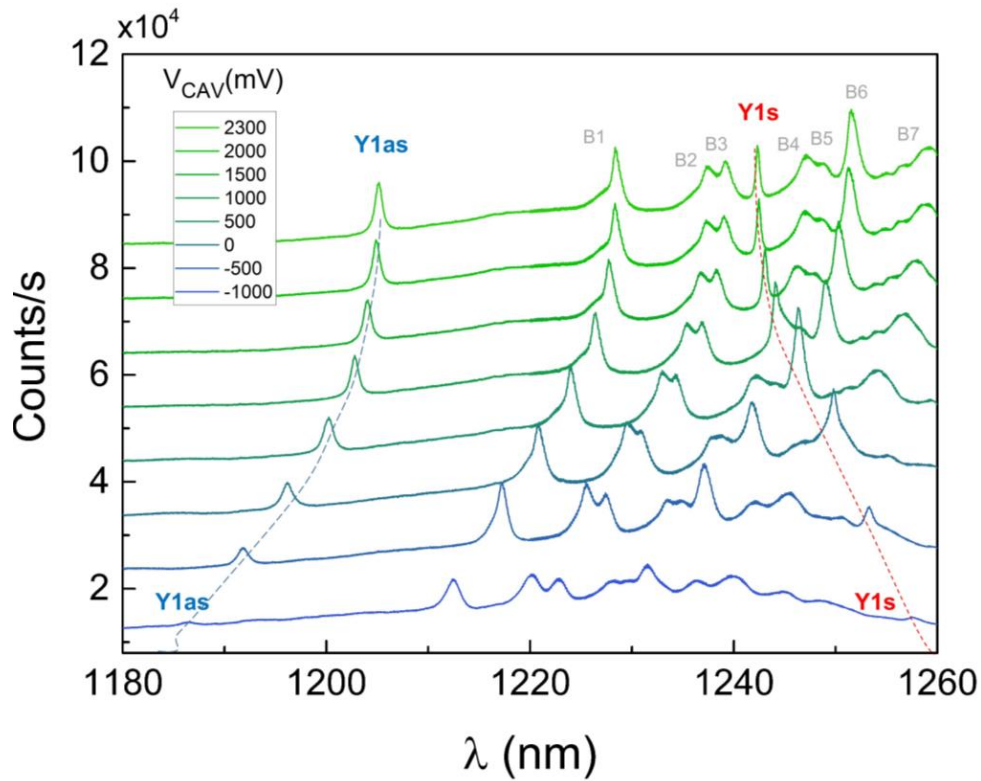

**Figure 2.** Electro-luminescence spectra of the photonic crystal device acquired at several cavity voltages (VCAV) with a voltage applied to the QD diode VQD = 3.5 V. The fundamental S (AS) mode shifts towards the blue (red) when the cavity voltage is increased. Several AS band edge modes (B1, ... B7) also appear in the spectrum. A maximum electromechanical tuning of more than 15.4 nm is here demonstrated. A constant offset (10kHz) between several measurements is introduced for clarity. The red and blue dashed lines are guides for the eye

Figure 2 shows the µEL signal of the cavity emission collected when the QD-diode is operated in forward bias ($V_{QD}$ = 3.5 V) and the voltage across the cavity-diode is swept from -1.0V to 2.3V. The spectra are characterized by several peaks that can be associated with the resonances of an L3 cavity by comparison with the local density of states calculated using 3D Finite-Element-Method (FEM) simulations. We can discriminate the vertical symmetry of the modes from the sign of the frequency tuning, since the S (AS) mode move towards higher (lower) wavelengths if the cavity actuation voltage is decreased. Indeed, the built-in charge on the actuation junction is increased when the applied voltage is decreased from positive to negative, corresponding to an increased attractive force and decreased membrane spacing. In this tuning experiment, all the modes manifest an AS character except the mode labelled as Y1s, identified as fundamental S mode, which shifts from 1242.3 to 1257.7 nm. The mode at the lowest wavelength is identified as fundamental antisymmetric mode, Y1as. Its wavelength can be reversibly varied over almost 20 nm from $\lambda_{-1V}$ = 1186.5 nm to $\lambda_{2.3V}$ = 1205.1 nm. Due to the high initial wavelength splitting of these two modes (37.2 nm), Y1s falls in the AS dielectric band where a series of AS band-edge modes is ascertained (B1, ..., B7). Notice that the intensity of the $\mu$EL signal decreases for negative values of $V_{CAV}$ due to the presence of a crosstalk between the two diodes that effectively reduces the measured injection current from 2.3 mA to 1.7 mA at $V_{CAV}$=-1V. Besides, the high current necessary to excite the PhC modes is attributed to the recombination close to the p-contact (current-crowding), which effectively reduces the current reaching the cavity. This induces a heating effect that results in a redshift of both the AS and S modes. As a consequence, the total tuning range of the AS mode is slightly larger compared to the S mode, and from their semi-difference we can extrapolate a thermal tuning range of 1.6 nm. Further optimization of the bridge geometry along with the use of lateral doping profiles [28] may bring this device into the lasing regime. Combining FEM simulations and the experimental S-AS splitting corrected with the thermal shift, we estimate a reduction of the inter-membrane distance from 200 nm to 145 nm in the actuation range investigated in these experiments.

In order to improve the vertical out-coupling efficiency, which is crucial for single-photon experiments, in the following we employed a device featuring a modified H0 cavity, whose far-field pattern approximately matches the Gaussian profile of the objective. When the injection current is set into the $\mu$A range by operating the QD-diode just above its threshold ($V_{th}$=1.2V), single dots lines emerge in the spectrum. Figure 3a shows color-coded $\mu$EL spectra acquired varying the voltage across the QD-diode, while setting $V_{CAV}$ = 2.2$V$. The variation in the applied QD-voltage introduces a blue-shift of the excitonic lines, attributed

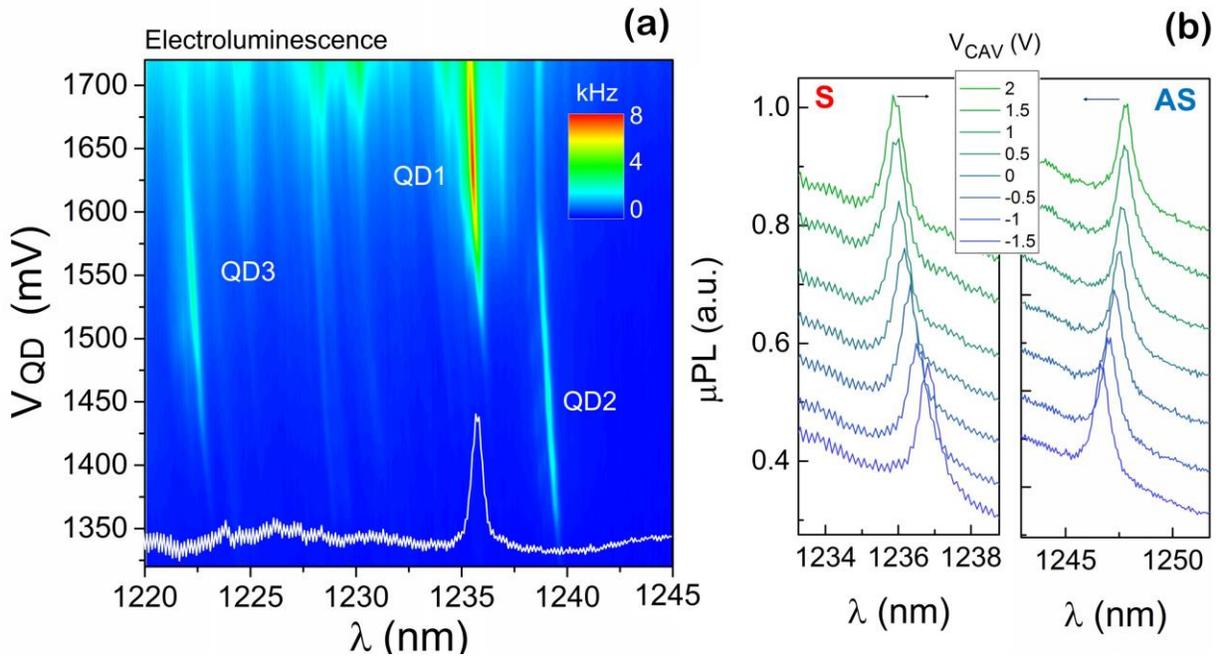

**Figure 3**. (a) Color-coded micro-electroluminescence spectra collected while varying the voltage across the QD (VQD). The bottom white line represents the µPL spectrum of the cavity mode (b) µPL spectra of the resonant modes when the cavity is actuated

to the quantum-confined Stark Effect. This effect appears also in forward bias, since the increase of the QD-voltage corresponds to a decrease in the modulus of the vertical field acting on the emitters. For the cavity actuation voltage chosen for this experiment, the excitonic line QD1 comes in resonance with the cavity mode identified from a $\mu$PL spectrum (white line) at a QD bias $V_{QD}$=1630 mV. A ten-fold enhancement of its electroluminescence signal is observed in this situation, compared to the case when it is detuned by 0.65 nm. Besides the single exciton line, emission within the cavity linewidth is also visible in the off-resonant µEL spectra, which is attributed to phonon-mediated feeding mechanisms [35]. Other excitonic lines (QD2, QD3) experience a similar Stark-shift, but do not cross the cavity resonance. We tentatively attribute these lines to other excitons, in the same or different dot, spectrally decoupled from the cavity mode. The cavity mode employed here, originating from the second-order mode of the H0 design, shows a quality factor $Q$ = 2270 and is characterized by a symmetric character as shown in the µPL cavity tuning spectra reported in Fig. 3b left panel.

Lastly, we investigate the non-classical nature of photons emitted by the cavity-enhanced QD1 line at $V_{QD}$=1630 mV by performing auto-correlation experiments. To this aim, we employed a fiber-based Hanbury Brown and Twiss (HBT) interferometer composed of a 50/50 beam splitter and two superconductive single-photon detectors [36](Efficiency ~ 45%, Dark Counts ~ 30Hz) interfaced to a correlation card (PicoHarp 300). Additionally, a fiber-coupled band-pass tuneable filter (FWHM=0.5 nm) is used to isolate the QD1 single line emission among the others (Fig. 4b).

Figure 4a shows the normalized auto-correlation histogram built from the raw coincidences of the two detectors as a function of time delay between the detection events ($\tau$). A clear dip, well below 0.5, is observed in the coincidences at zero-time delay, which is the clear signature of anti-bunching in the emission from a single quantum emitter. The raw value at zero-time delay is $g^{(2)}_{raw}(0) = 0.11$. By fitting the anti-bunching curve with the function $g^{(2)}(\tau) = 1 - A \exp(|t|/\tau_t)$ (red curve) we can extract the zero-delay second order coherence function $g^{(2)}(0) = 1 - A = 0.13 \pm 0.09$ and the decay time associated with this transition $\tau_t$ = (420 ± 20) ps, assuming the QD is in the low-excitation regime [37]. (the error bars are derived from the standard deviations of the fit). The non-zero value of $g^{(2)}(0)$ is attributed to the residual background from the cavity mode emission, which contributes by 16% to the total µEL signal integrated in the filter range. The low detector jitter (~50 ps) provides enough resolution to temporally resolve the dip without need for deconvolution with the instrument response function.

The fact that the time constant associated to the antibunching dip is shorter than the QD lifetime in the bulk (see discussion below) is indicative of spontaneous emission enhancement in the cavity mode. While the assumption of low excitation cannot be proved in this particular device, as the injection rate and the QD-cavity detuning cannot be controlled separately, the strong increase in count rate as the QD line crosses the mode (Fig. 3(a)) provides an additional indication of the cavity-induced enhancement of the emission rate. In general, the decay dynamics of an emitter coupled to a cavity in the presence of a static electric field can be decomposed in several decay channels $\tau^{-1} = \tau_{PhC}^{-1} + \tau_{leaky}^{-1} + \tau_{tun}^{-1}$, where $\tau_{PhC}^{-1}$ is the resonant decay into the mode, $\tau_{leaky}^{-1}$ represents the decay channel in the leaky modes and is negligible in photonic crystal cavities ($\tau_{leaky}$ = 3−6 ns [38]), while $\tau_{tun}^{-1}$ is associated with the tunnelling rate out of the QD, which strongly depends on the applied electric field. In order to measure $\tau_{tun}^{-1}$ we performed µPL time-resolved measurements by exciting the dot ensemble in a bulk region with a 750nm-pulsed laser (instrument response function = 90ps) while varying the voltage applied to the dots $V_{QD}$.

Figure 4c shows the fast decay component of the dot emission as a function of $V_{QD}$.

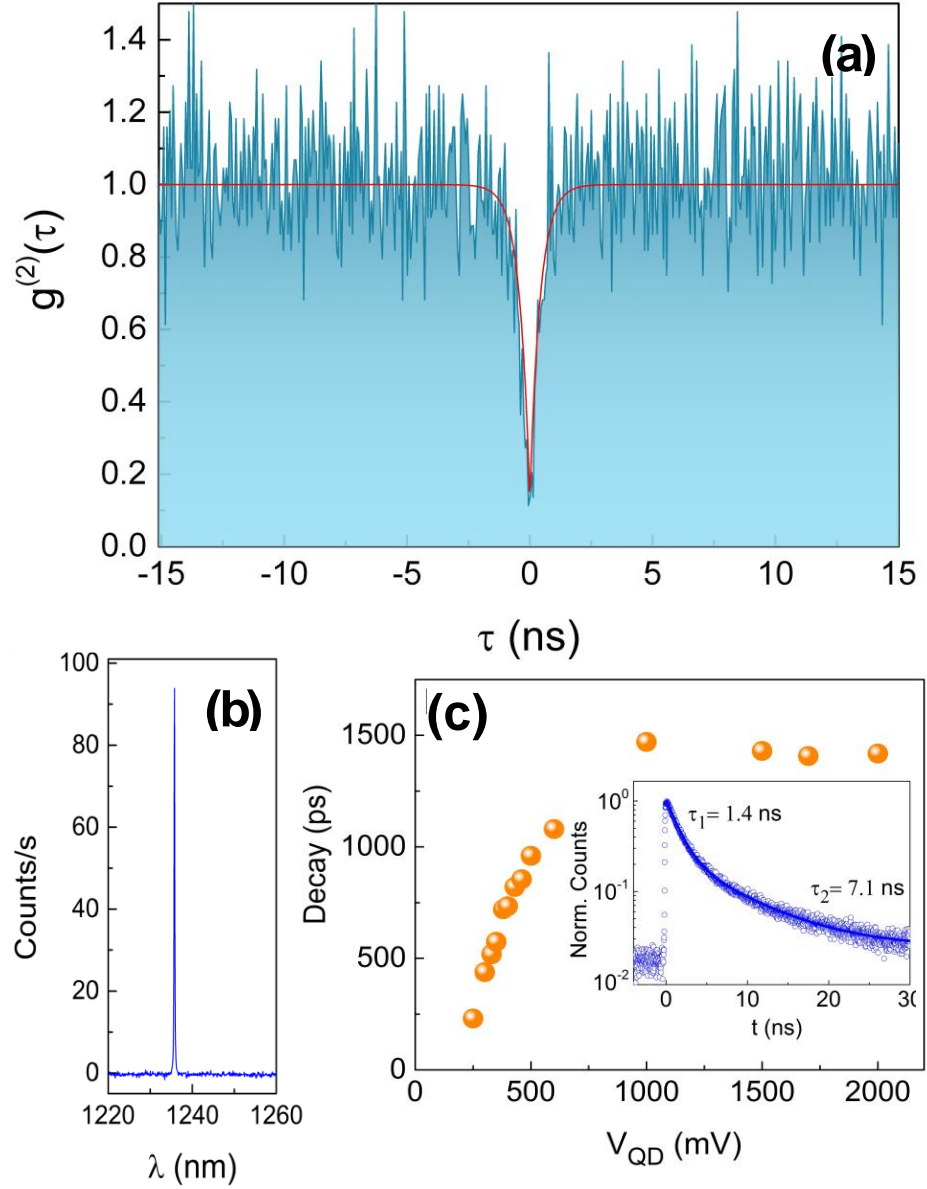

**Figure 4**. (a) Auto-correlation histogram of the dot QD1 at $V_{QD}$=1630 mV fulfilling the resonance condition with the cavity mode. (b) Bright single line emission after filtering. (c) Decay constants extracted from time-resolved (TR) µPL experiments of the QD ensemble as a function of the voltage applied to the dot layer. The inset show the TR µPL decay for $V_{QD}$=1.5V

It is evident that for $V_{QD}$ < 1.2V the decay time decreases when the field ($V_{QD}$) is increased (decreased). This represents a clear evidence that in this voltage range the decay is dominated by a non-radiative process associated with the tunnelling mechanism. Instead, when the QD-diode is operated at $V_{QD}$ > 1.2V, the decay time does not depend on the applied voltage. This value corresponds to the radiative time at zero field $\tau_{bulk}$ = 1.45 ns [38, 32]. A slower decay component is also visible in the time-resolved data (Figure 4c inset) and is attributed to the presence of a dark exciton repopulating the bright transition. This second decay constant shows a similar dependence on the applied voltage.

These experiments prove that $\tau_{tun}^{-1}$ is negligible for the injection experiments shown in Figure 3a and, consequently, that the decay dynamics is dominated by radiative emission in these conditions. The operation speed of the current device is limited to few MHz both by the measured series resistance of the QD-diode ($R = 1.5\text{k}\Omega$) and its estimated capacitance (C~15pF). While this has prevented the operation in pulsed mode for the reported device, it can be easily overcome by improvements in the technology. The contact resistance can be reduced below 200Ω by optimizing the metals alloy employed for the contacts, while the parasitic capacitance can be decreased to below ~1pF with a reduced contact area, as demonstrated for a single-membrane PhC diode [39]. This further optimization will provide a cut-off frequency of the QD-diode above few GHz and therefore will enable the operation in the pulsed regime, allowing for the direct measurement of the lifetime and the operation as on-demand single-photon source.

**CONCLUSIONS**

In summary, we reported the non-classical light emission from electrically-injected quantum dots in a photonic crystal cavity. The full electrical control of the energy position of cavities and quantum dots, needed to build scalable Purcell-enhanced sources, has been achieved by the combination of electrostatic actuation and Stark tuning. The results presented here can be easily extended to more complex photonic crystal environments, such as slow-light waveguides or chiral structures [40]. While the tuning range of QD lines in this configuration is limited due to the coupling between tuning voltage and injection current in the QD junction, it could be extended by implementing additional strain tuning structures [41]. Importantly, the integration with linear components such as ridge waveguides and phase modulators will pave the way to on-demand single-photon experiments on a single architecture.

**ACKNOWLEDGEMENTS**

This research was financially supported by the Dutch Technology Foundation STW, Applied Science Division of NWO, the Technology Program of the Ministry of Economic Affairs under projects Nos. 10380 and 12662 and by NanoNextNL, a micro and nanotechnology program of the Dutch Ministry of Economic Affairs, Agriculture and Innovation (EL&I) and 130 partners. This work is also part of the research programme of the Foundation for Fundamental Research on Matter (FOM), which is financially supported by the Netherlands Organisation for Scientific Research (NWO).